\documentclass[final,3p,times,onecolumn,preprint]{elsarticle}
\usepackage{lineno,hyperref}

\usepackage{amsmath}
\usepackage{subfig}
\usepackage{graphicx,epsfig,wrapfig}

\usepackage{amsthm,mathrsfs}
\usepackage{makeidx}
\usepackage{amsfonts}
\usepackage{amssymb}
\usepackage{amsmath}
\usepackage{mathtools}
\usepackage{fixmath}
\usepackage{amsbsy}
\usepackage[capitalize]{cleveref}
\usepackage{mathrsfs}
\usepackage{slashed}
\usepackage{bm}
\usepackage{multirow}
\usepackage{booktabs}

\usepackage{color}
\usepackage[normalem]{ulem}

\definecolor{myGreen}{rgb}{0.2,0.72,0.2}
\renewcommand\sout{\bgroup \color[rgb]{0.55,0.00,0.99} \ULdepth=-.5ex \ULset}

\usepackage{simplewick}

\usepackage{multirow}
\usepackage{braket}

\newcommand{\ta}{\left(}
\newcommand{\qa}{\left[}

\newcommand{\tc}{\right)}
\newcommand{\qc}{\right]}

\renewcommand{\[}{\begin{equation}}
\renewcommand{\]}{\end{equation}}

\makeatletter
\newcommand*\dotp{\mathpalette\dotp@{.5}}
\newcommand*\dotp@[2]{\mathbin{\vcenter{\hbox{\scalebox{#2}{$\m@th#1\bullet$}}}}}
\makeatother

\begin{document}
\begin{frontmatter}

\title{Analytical derivatives of Neural Networks}
\author[mymainaddress,mysecondaryaddress]{Simone Rodini\corref{mycorrespondingauthor}}\fnref{fn1}
\cortext[mycorrespondingauthor]{Corresponding author}
\ead{simone.rodini@unipv.it}
\fntext[fn1]{\textit{Phone number:} +39 0382987447}
\address[mymainaddress]{Dipartimento di Fisica, Universit\`a degli Studi di Pavia, I-27100 Pavia, Italy}
\address[mysecondaryaddress]{Istituto Nazionale di Fisica Nucleare, Sezione di Pavia, I-27100 Pavia, Italy}

\begin{abstract}
We propose a simple recursive algorithm that allows the computation of the first- and second-order derivatives with respect to the inputs of an arbitrary deep feed forward neural network (DFNN). The algorithm naturally incorporates the derivatives with respect to the network parameters. To test the algorithm, we apply it to the study of the quantum mechanical variational problem for few cases of simple potentials, modeling the ground-state wave function in terms of a DFNN.
\end{abstract}

\date{\today}


\end{frontmatter}

\newcommand{\yd}{\overset{\dotp}{y}}
\newcommand{\ydd}{\overset{\dotp\dotp}{y}}
\newcommand{\yddd}{\overset{\dotp\dotp\dotp}{y}}

\section{Introduction}
In recent years, physical informed neural networks (PINNs) have attracted increasing interest. The overall idea is to utilize a deep feed forward neural network (DFNN) to approximate the solution to a given differential equation that describes a physical system or phenomena. The advantage of using DFNNs, compared to more traditional methods or to different types of trial solutions, is the possibility of virtually studying all differential equations with an unified perspective, thanks to the fact that the DFNNs are able to approximate any reasonable functional form, see \cite{Cybenko:1989}. For a  discussion about the advantages and disadvantages of using DFNN over more traditional methods see \cite{Avrutskiy_2020,dockhorn2019discussion}.

In this work, we are going to use the fundamental concepts of PINNs \cite{LEE1990110,raissi2017physics1,raissi2017physics2,Lagaris_1998,Lagaris_2000,Lagaris_1997,Guidetti:2021xvb} to construct the ground-state solutions for different types of quantum mechanical potentials. One could use two different approaches to solve the problem. 
The first one is a pure PINN approach, in which one tries to solve the eigenvalue differential equation by modeling the ground-state wave function with a DFNN and  by computing the approximation of the energy at each iteration as the expectation value of the Hamiltonian on the approximate state, see \cite{Lagaris_1997}. This approach is rather slow, since for each iteration one needs not only to evaluate the differential equation in a sufficiently large number of points but also to estimate the integral (in general multi-dimensional) associated with the expectation value of the Hamiltonian. This basically doubles the computational time with respect to a ``standard'' PINN in which the only unknown is the function solution.  

Therefore, we relied on a second approach, which uses the variational approach to quantum mechanics. The advantage of the variational approach consists in removing the evaluation of the differential equation associated to the eigenvalue problem, leaving only the computation of the Hamiltonian expectation value. 
From the equivalence of the variational approach to the standard formulation of the eigenvalue problem, we can construct not only the ground-state energy, but also the wave function. The investigation of the variational approach to PINN has been started in \cite{kharazmi2019variational} for shallow networks. A discussion for an approach to variational problems using deep networks can be found in \cite{Wu_2017} The application of the variational approach to quantum mechanical problems using shallow networks was already investigated in \cite{nakanishi2000,sugawara2001,MARIM2003159}.
 
In order to carry out the program of solving the eigenvalue problem, we wanted to avoid automatic differentiation of the network structure, since, albeit faster than numerical differentiation, it is still slower than an analytical approach. 
We therefore derived the analytical expressions for the first- and second-order derivatives of a generic DFNN, along with all the relevant derivatives with respect to the parameters of the DFNN. This analytical approach works best for small to mid-sized input spaces. For large input spaces, one should consider to either modify the algorithm to compute only the relevant derivatives, or switch to a Monte Carlo approach to compute higher-order derivatives, see \cite{Sirignano_2018}.
However, for appropriate input-space sizes, the analytical approach to the derivatives should be the preferential choice.

The work is organized as follows. In section \ref{SecAnalytDer}, we present the algorithm to compute the analytical derivatives of a generic DFNN. In section \ref{sectQM}, we give a general overview of the variational method in quantum mechanics, outlying the assumptions upon which our application relies. In section \ref{SecExamples}, we analyze a few mono-dimensional potentials to demonstrate the applicability of the algorithm. Finally, in section \ref{SecGeneralization}, we outcurve the possible extensions to multi-dimensional potentials and we analyze the evaluation-time of our implementation of the algorithm. 
In section \ref{SecConcl}, we present a summary of the work.

\section{Analytical derivatives of Neural Networks}
\label{SecAnalytDer}
This section will be devoted to the construction of derivatives of a generic DFNN. We will generally follow the notation introduced in \cite{AbdulKhalek:2020uza}, modifying it when needed.

We suppose to have a DFNN with $L$ layers, indexed from $0$ to $L$, where the $0$-th layer corresponds to the inputs. The number of nodes in the layer $l$ is denoted by $N_l$, with the understanding that $N_0$ represent the number of inputs to the network and $N_L$ the number of outputs.
The output of the network is computed as:
\[
O_k = \sigma_L\ta b_{k}^{(L)} + \sum_{j=1}^{N_{L-1}}\omega^{(L)}_{kj}y_j^{(L-1)}\tc \equiv y_k^{(L)}, \quad k=1,...,N_L, 
\]
with the recurrence relation
\[
y_k^{(l)} = \sigma_l\ta b_{k}^{(l)} + \sum_{j=1}^{N_{l-1}}\omega^{(l)}_{kj}y_j^{(l-1)}\tc,
\quad \quad y_k^{(0)} = \sigma_0\ta \xi_k\tc, 
\label{input_bias}
\]
where $\bm{\xi}\in \Omega \subset \mathbb{R}^{N_0}$ is the input of the neural network. 
The functions $\sigma_l$ are usually referred to as ``activation'' functions. Different functional forms are used in the literature, among which one of the most famous is the sigmoid function $\sigma(x) = 1/(1+e^{-x})$. Throughout this work we are going to use as the activation function for the hidden layers the hyperbolic tangent function $\sigma_l(x) =\tanh(x)$ if $l\neq 0,L$, which offers the advantage of positive and negative values for the activated hidden nodes. For the input and output layer we are going to use instead the identity function $\sigma_{0,L}(x) = \text{id}(x) = x$. However, in problems in which the input variable $\bm{\xi}$ can have components $\xi_a$ with relatively large differences in absolute values, the activation function $\sigma_0$ can be used to equalize the ranges of $\xi_a$.

Following Ref.~\cite{AbdulKhalek:2020uza}, the first derivatives of the network with respect to the parameters are given by the following relation: 
\[
\frac{\partial O_k}{\partial \pi^{(l)}} = \Sigma^{(l)}_{ki}\yd_i^{(l)}\begin{cases}
1 & \text{ if } \pi^{(l)} = b_i^{(l)}\\
y_j^{(l-1)} & \text{ if } \pi^{(l)} = \omega_{ij}^{(l)}\\
\end{cases},
\label{derPi}
\]
where we defined $\yd_i^{(l)}$ as follows (see also Eq.~\eqref{dotDef}):
\[
\yd_i^{(l)} = \frac{\partial \sigma_l(z)}{\partial z}\Bigg|_{z=\ta b_{i}^{(l)} + \sum_{j=1}^{N_{l-1}}\omega^{(l)}_{ij}y_j^{(l-1)}\tc}.
\]
In the following we are going also to reserve the symbol $\partial /\partial \xi_a$ to denote the derivatives with respect to the inputs.

Hereafter, in any system of equation as in \eqref{derPi}, the top row is implicitly assumed to refer to the case $\pi^{(l)} = b_i^{(l)}$ and the bottom row to the case  $\pi^{(l)} = \omega_{ij}^{(l)}$.
The matrix $\Sigma^{(l)}_{ki}$ can be computed from the following recursive relation:
\[
\Sigma^{(l)}_{ki} = \sum_{j_{l+1}=1}^{N_{l+1}} \Sigma^{(l+1)}_{k j_{l+1}}S^{(l+1)}_{j_{l+1} i}, \quad \quad S^{(l)}_{ji} = \yd_j^{(l)}\omega_{ji}^{(l)}, \quad \quad \Sigma^{(L)}_{ki} = \delta_{ki}.
\]

For $l=0$, Eq.~\eqref{derPi} gives the  first derivatives of the network with respect to the inputs. We now need to obtain an expression for the double derivatives of the network with respect to the inputs and the parameters. 

We have form Eq.~\eqref{derPi}: 
\[
\frac{\partial}{\partial \xi_a}\frac{\partial O_k}{\partial \pi^{(l)}} = \ta \frac{\partial  \Sigma^{(l)}_{ki}}{\partial \xi_a} \yd_i^{(l)} + \frac{\partial  \yd_i^{(l)} }{\partial \xi_a} \Sigma^{(l)}_{ki}\tc \begin{cases}
1 \\
y_j^{(l-1)} 
\end{cases}  + 
 \Sigma^{(l)}_{ki} \yd_i^{(l)}\begin{cases}
0 \\
\frac{\partial  y_j^{(l-1)} }{\partial \xi_a}
\end{cases}.
\label{derPiderXi}
\]
The fundamental building block of the above relation is the following (recursive) identity
\[
\frac{\partial \!\phantom{y}_{ (n)}\yd\ _j^{(l)} }{\partial \xi_a} = \!\phantom{y}_{ (n+1)}\yd\ _j^{(l)} \sum_{i=1}^{N_{l-1}} \omega_{ji}^{(l)} \frac{\partial y_i^{(l-1)} }{\partial \xi_a},
\label{fund_der1}
\]
with seed
\[
\frac{\partial y_j^{(1)} }{\partial \xi_a} = \omega_{ja}^{(1)}\overset{\dotp}{\sigma}_0(\xi_a), \quad\quad \frac{\partial y_b^{(0)} }{\partial \xi_a} = \delta_{ab}\overset{\dotp}{\sigma}_0(\xi_a),
\]
where $\!\phantom{y}_{ (n)}\yd\ _j^{(l)}$ stands for the $n$-th derivative of $y_j^{(l-1)}$, i.e.
\[
\!\phantom{y}_{ (n)}\yd\ _j^{(l)} = \frac{\partial^n \sigma_l(z)}{\partial z^n}\Bigg|_{z=\ta b_{k}^{(l)} + \sum_{j=1}^{N_{l-1}}\omega^{(l)}_{kj}y_j^{(l-1)}\tc}.
\label{dotDef}
\]
Using Eq.~\eqref{fund_der1}, we can obtain the derivative:
\begin{align}
&\frac{\partial  \Sigma^{(l)}_{ki}}{\partial \xi_a} = \frac{\partial }{\partial \xi_a}\sum_{\substack{\{j_f\}\\ f=L,...,l+1}} S^{(L)}_{k\ j_{L-1}} ... \ S^{(l+1)}_{j_{l+1}\ i}\notag \\
&=  \frac{\partial }{\partial \xi_a}\sum_{\substack{\{j_f\}\\ f=L,...,l+1}} \delta_{j_L k}\delta_{j_l i}\prod_{n=L}^{l+1}S^{(n)}_{j_n\: j_{n-1}}=  \sum_{\substack{\{j_f\}\\ f=L,...,l+1}} \delta_{j_L k}\delta_{j_l i} \sum_{m=L}^{l+1}\frac{\partial S^{(m)}_{j_m\:j_{m-1}}}{\partial \xi_a}\prod_{\substack{n=L\\ n\neq m}}^{l+1}S^{(n)}_{j_n\: j_{n-1}},
\label{sigma_der1}
\end{align}
with 
\[
\frac{\partial S^{(l)}_{ji}}{\partial \xi_a} = \frac{\partial \yd_j^{(l)}}{\partial \xi_a}\omega_{ji}^{(l)}.
\]

Eq.~\eqref{sigma_der1} can be recursively computed starting from the output layer by virtue of the following identities:
\begin{align}
\Sigma_{kj}^{(l)} &= \sum_{i=1}^{N_{l+1}} \Sigma^{(l+1)}_{ki} S^{(l+1)}_{ij} = \sum_{i=1}^{N_{l+1}} \Sigma^{(l+1)}_{ki} \omega^{(l+1)}_{ij} \ \overset{\dotp}{y}_i^{(l+1)},\\
\frac{\partial  \Sigma^{(l)}_{kj}}{\partial \xi_a}&= \sum_{i=1}^{N_{l+1}} \Bigg\{\frac{\partial  \Sigma^{(l+1)}_{ki}}{\partial \xi_a} \omega^{(l+1)}_{ij} \ \overset{\dotp}{y}_i^{(l+1)} +  \Sigma^{(l+1)}_{ki} \omega^{(l+1)}_{ij} \ \frac{\partial \ \overset{\dotp}{y}_i^{(l+1)}}{\partial \xi_a}\Bigg\} ,
\label{sigma_der1_rec}
\end{align}
with seed
\[
\frac{\partial  \Sigma^{(L)}_{kj}}{\partial \xi_a} = 0.
\]

Eq.~\eqref{derPiderXi} provides the full parameter gradient of the derivative of the network with respect to the inputs. Moreover, as well as Eq.~\eqref{derPi} for $l=0$ gives the derivatives of the network with respect to the inputs, Eq.~\eqref{derPiderXi} for $l=0$ gives the hessian of the network with respect to the inputs, which could be used in hessian-based minimization algorithms.
To obtain the hessian of the network with respect to the parameters, one simply replaces $\xi_a$ by $\tilde{\pi}^{l'}$ and eliminates the appropriate derivatives as, for example,
\[
\frac{\partial \ \overset{\dotp}{y}_i^{(l)}}{\partial \tilde{\pi}^{l'}} = 0, \quad \text{ if } l'>l.
\]

The only remaining step  is the computation of the second derivatives with respect to the inputs, for which we find:
\begin{align}
& \frac{\partial}{\partial \xi_b}\frac{\partial}{\partial \xi_a}\frac{\partial O_k}{\partial \pi^{(l)}} \notag\\
&=  \frac{\partial^2  \Sigma^{(l)}_{ki}}{\partial \xi_a\partial \xi_b} \yd_i^{(l)}  \begin{cases}
1 \\
y_j^{(l-1)} 
\end{cases} +   \frac{\partial  \Sigma^{(l)}_{ki}}{\partial \xi_a} \frac{\partial \yd_i^{(l)}}{\partial \xi_b}  \begin{cases}
1 \\
y_j^{(l-1)} 
\end{cases} + \frac{\partial  \Sigma^{(l)}_{ki}}{\partial \xi_a} \yd_i^{(l)}  \begin{cases}
0 \\
\frac{\partial y_j^{(l-1)} }{\partial \xi_b}
\end{cases} \notag\\
&+ \frac{\partial^2  \yd_i^{(l)} }{\partial \xi_a\partial \xi_b} \Sigma^{(l)}_{ki} \begin{cases}
1 \\
y_j^{(l-1)} 
\end{cases} + \frac{\partial  \yd_i^{(l)} }{\partial \xi_a} \frac{\partial  \Sigma^{(l)}_{ki}}{\partial \xi_b} \begin{cases}
1 \\
y_j^{(l-1)} 
\end{cases} + \frac{\partial  \yd_i^{(l)} }{\partial \xi_a} \Sigma^{(l)}_{ki} \begin{cases}
0 \\
\frac{\partial  y_j^{(l-1)} }{\partial \xi_b} 
\end{cases} \notag\\
&+ \frac{\partial  \Sigma^{(l)}_{ki}}{\partial \xi_b} \yd_i^{(l)}\begin{cases}
0 \\
\frac{\partial  y_j^{(l-1)} }{\partial \xi_a}
\end{cases} 
+ \Sigma^{(l)}_{ki} \frac{\partial  \yd_i^{(l)} }{\partial \xi_b}\begin{cases}
0 \\
\frac{\partial  y_j^{(l-1)} }{\partial \xi_a}
\end{cases} 
+ \Sigma^{(l)}_{ki} \yd_i^{(l)}\begin{cases}
0 \\
\frac{\partial^2  y_j^{(l-1)} }{\partial \xi_a\partial \xi_b}
\end{cases}.
\label{derPiderXiderXi}
\end{align}
We now need the recursive relations for the double derivatives.
From Eq.~\eqref{fund_der1}, we have:

\begin{align}
\frac{\partial^2 \!\phantom{y}_{ (n)}\yd\ _j^{(l)} }{\partial \xi_a\partial \xi_b} &=\frac{\partial}{\partial \xi_b}\frac{\partial \!\phantom{y}_{ (n)}\yd\ _j^{(l)} }{\partial \xi_a} = \frac{\partial}{\partial \xi_b} \ta \!\phantom{y}_{ (n+1)}\yd\ _j^{(l)} \sum_{i=1}^{N_{l-1}} \omega_{ji}^{(l)} \frac{\partial y_i^{(l-1)} }{\partial \xi_a}\tc \notag \\
& = \frac{\partial \!\phantom{y}_{ (n+1)}\yd\ _j^{(l)}}{\partial \xi_b}   \sum_{i=1}^{N_{l-1}} \omega_{ji}^{(l)} \frac{\partial y_i^{(l-1)} }{\partial \xi_a} + \!\phantom{y}_{ (n+1)}\yd\ _j^{(l)} \sum_{i=1}^{N_{l-1}} \omega_{ji}^{(l)} \frac{\partial^2 y_i^{(l-1)} }{\partial \xi_a\partial \xi_b} \notag \\
& =  \!\phantom{y}_{ (n+2)}\yd\ _j^{(l)}  \ta \sum_{i=1}^{N_{l-1}} \omega_{ji}^{(l)} \frac{\partial y_i^{(l-1)} }{\partial \xi_a}\tc \ta \sum_{i=1}^{N_{l-1}} \omega_{ji}^{(l)} \frac{\partial y_i^{(l-1)} }{\partial \xi_b}\tc + \!\phantom{y}_{ (n+1)}\yd\ _j^{(l)} \sum_{i=1}^{N_{l-1}} \omega_{ji}^{(l)} \frac{\partial^2 y_i^{(l-1)} }{\partial \xi_a\partial \xi_b}.
\end{align}
From Eq.~\eqref{sigma_der1}, we can derive the following relation
\begin{align}
&\frac{\partial^2  \Sigma^{(l)}_{ki}}{\partial \xi_a\partial \xi_b} \notag \\
&=  \sum_{\substack{\{j_i\} \\ i=L,...,l+1}} \delta_{j_L k}\delta_{j_l i} \qa \sum_{m=L}^{l+1}\frac{\partial^2 S^{(m)}_{j_m\:j_{m-1}}}{\partial \xi_a\partial \xi_b}\prod_{\substack{n=L\\n\neq m}}^{l+1}S^{(n)}_{j_n\: j_{n-1}}  + \sum_{\substack{m,g=L\\ m\neq g}}^{l+1}\frac{\partial S^{(m)}_{j_m\:j_{m-1}}}{\partial \xi_a}\frac{\partial S^{(g)}_{j_g\:j_{g-1}}}{\partial \xi_b}\prod_{\substack{n=L\\ n\neq m , g}}^{l+1}S^{(n)}_{j_n\: j_{n-1}}\qc,
\end{align}
with 
\[
\frac{\partial^2 S^{(l)}_{ji}}{\partial \xi_a\partial \xi_b} = \frac{\partial^2 \yd_j^{(l)}}{\partial \xi_a\partial \xi_b}\omega_{ji}^{(l)}.
\]
By exploiting the relation in Eq.~\eqref{sigma_der1_rec}, we can recursively construct the second derivatives of $\Sigma^{(l)}$ starting from the output layer:
\begin{align}
\frac{\partial^2  \Sigma^{(l)}_{kj}}{\partial \xi_a\partial \xi_b}&= \sum_i \Bigg\{ \frac{\partial^2  \Sigma^{(l+1)}_{ki}}{\partial \xi_a\partial \xi_b} \omega^{(l+1)}_{ij} \ \overset{\dotp}{y}_i^{(l+1)} + \Sigma^{(l+1)}_{ki} \omega^{(l+1)}_{ij} \ \frac{\partial^2 \ \overset{\dotp}{y}_i^{(l+1)}}{\partial \xi_a\partial \xi_b} \notag \\
& + \frac{\partial  \Sigma^{(l+1)}_{ki}}{\partial \xi_a} \omega^{(l+1)}_{ij} \frac{\partial \ \overset{\dotp}{y}_i^{(l+1)}}{\partial \xi_b} +  \frac{\partial  \Sigma^{(l+1)}_{ki}}{\partial \xi_b} \omega^{(l+1)}_{ij} \frac{\partial \ \overset{\dotp}{y}_i^{(l+1)}}{\partial \xi_a}\Bigg\},
\end{align}
with seed
\[
\frac{\partial^2  \Sigma^{(L)}_{kj}}{\partial \xi_a\partial \xi_b} = 0.
\]

This completes the derivation. The three equations \eqref{derPi}, \eqref{derPiderXi} and \eqref{derPiderXiderXi} allow one to analytically compute any expression of the following form
\[
\mathcal{O}\ta \nabla_{ab}^2\bm{N}, \nabla_a \bm{N}, \bm{N} \tc, \quad a,b = 1,..., N_0, 
\]
and its gradient with respect to the DFNN parameters $\pi^{(l)}$
\[
\nabla_{\pi^{(l)}}\mathcal{O}\ta \nabla_{ab}^2\bm{N}, \nabla_a \bm{N}, \bm{N} \tc, \quad a,b = 1,..., N_0.
\]

\section{ground-state determination in Quantum Mechanics}
\label{sectQM}
In the following two sections, we are going to showcase the algorithm of section \ref{SecAnalytDer} using the common task in Quantum Mechanics (QM) of solving the eigenvalue equation of the Hamitlonian for the ground-state in the case of a generic potential.  
In this section we briefly review the variational method for the computation of the ground-state of a generic system and we are going to lay out few fundamental assumptions. In the next section we present the results.

The generic eigenvalue problem is:
\[
H\ket{\psi_0} =\ta K + V\tc \ket{\psi_0}= E_0\ket{\psi_0},\quad \quad \braket{\psi_0|\psi_0} = 1,
\label{SchrEigenEq}
\]
where $K$ is the kinetic energy part of the Hamiltonian, $V$ the potential and $\ket{\psi_0}$ the ground state. We are using the Dirac bra-ket notation to denote a generic state in the appropriate Hilbert space $\mathcal{H}$. 
The analytical solution of the problem is, in general, very challenging. Therefore, one has to resort to different approximation methods 
 that provide  suitable numerical solutions. Here we are going to focus on the so-called variational methods. Being $\ket{\phi}$ a generic approximation of the true ground-state $\ket{\psi_0}$, the following statement holds (see, e.g. \cite{Sakurai:2011zz}):
\[
\bar{E} = \frac{\braket{\phi| H|\phi}}{\braket{\phi|\phi}} \ge \frac{\braket{\psi_0| H|\psi_0}}{\braket{\psi_0|\psi_0}} = E_0, \quad  \forall \ket{\phi}\in\mathcal{H},
\label{VariationalCond}
\]
where $\bar{E}$ is the energy functional.
Any particular implementation of the variational method establishes a set of rules to update the ground-state approximation $\ket{\phi}$ that  minimizes $\bar{E}$.  It can be proven that the set of conditions:
\[
\tilde{E} = \min_{\ket{\phi}\in \mathcal{H}}\bar{E}, \quad \quad \braket{\tilde{\phi}|\tilde{\phi}} = 1
\]
is equivalent to the eigenvalue problem  \eqref{SchrEigenEq}, i.e. $\tilde{E} = E_0$, $\ket{\tilde{\phi}} = \ket{\psi_0}$.

Our aim is to use a DFNN to approximate the ground-state for different potentials. For the sake of demonstration, we are going to  work in position-space representation, in which Eq.~\eqref{SchrEigenEq} takes the form (for $n$ spatial dimensions):
\[
\ta -\frac{\hbar^2}{2m}\nabla^2 + V(\bm{x})\tc \psi_0(\bm{x}) = E_0\psi_0(\bm{x}).
\]
Hereafter, we adopt a unit system in which $\hbar =1$ and we assume $m=1$.
The variational condition \eqref{VariationalCond} takes the form:
\[
\bar{E} = \ta\int_{\mathbb{R}^n} d\bm{x} \phi^*(\bm{x}) H\phi(\bm{x})\tc \bigg/ \ta\int_{\mathbb{R}^n} d\bm{x} \phi^*(\bm{x}) \phi(\bm{x})\tc.
\label{PosSpaceVarCond}
\]
We are going to assume the following \textit{essential} conditions:
\begin{enumerate}
\item \label{ess1} The ground-state wave function is sufficiently localized in position space, i.e. for any $1>\epsilon >0$ we can find a compact subregion $\Omega\subset \mathbb{R}^n$ such that 
\[
\int_{\Omega} d\bm{x} \phi^*(\bm{x}) \phi(\bm{x}) = 1-\epsilon.
\]
\item The ground-state eigenvalue problem has a solution $E_0$ such that $-\infty < E_0 <\infty$.
\item It exists a reliable algorithm to evaluate a discrete approximation of the integrals in Eq.~\eqref{PosSpaceVarCond}.
\end{enumerate}
These conditions  usually hold for a very large class of problems.
We are going to assume also the following working conditions (they are inessential, in the sense that the solution of the problem can be generalized by relaxing one or more of these conditions):
\begin{itemize}
\item The ground-state wave function is real, i.e. $\phi^*(\bm{x}) = \phi(\bm{x})$ (a complex wave function can always be modeled by a bi-output DFNN).
\item The problem is mono-dimensional, i.e. $\bm{x} = x\in \mathbb{R}$ (a multi-dimensional problem requires a meshed or Monte Carlo integration algorithm).
\item The ground-state energy is positive, i.e. $E_0>0$ (for a general potential that satisfies the essential conditions one can always shift the potential by a constant amount to ensure the validity of this condition).
\end{itemize}
As a final remark, we notice that,  for problems with an unknown solution, the goodness of the approximate solution $\{\bar{E},\phi(x)\}$ can be directly checked using Eq.~\eqref{SchrEigenEq}:
\[
\mathcal{G}(N) = \frac{1}{N}\sum_{i=1}^N \ta H\phi(x_i) - \bar{E}\phi(x_i)\tc^2.
\]
\section{Study cases and examples}
\label{SecExamples}
To minimize the energy functional \eqref{PosSpaceVarCond}, we choose to use ADAM minimizer \cite{Kingma:2014vow}, since it provides a good compromise between the ability of overcoming local minima and overall convergence. 
Throughout this section, all the considered DFNNs have hyperbolic tangent activation functions in the hidden layers and identity activation functions on the input and output layers. 
Under the conditions of section \ref{sectQM}, we use a discretization of the integrals by rectangular approximation with $O(10^4)$ points.

We study the following potentials:
\begin{itemize}
\item{Harmonic potential:}
\begin{align}
V(x) &= \frac{1}{2}x^2, \quad E_0 = \frac{1}{2}, \quad 
\psi_0(x) = e^{-\frac{x^2}{2}} \pi^{-\frac{1}{4}}.
\end{align}

A simple DFNN with six hidden layers with ten nodes each already gives excellent results:
\[
\bar{E}-E_0 <  10^{-6},\quad \quad |\braket{\phi|\phi} - 1| <  10^{-7}, \quad \quad \mathcal{G}(10^5) \simeq  10^{-6}.
\]
In figure \ref{harmonicPsi}  is shown the difference between the DFNN approximation and the exact solution as a function of $x$. The agreement is satisfactory, for both the absolute and difference. In the low- and high-$x$ region the relative difference grows, because of the vanishing of the wave function at the boundary. 
The integration region $\Omega$ assumed by the essential condition \ref{ess1} is chosen to be $\Omega = [-4,4]$.

\begin{figure}
\centering
\subfloat[]{\includegraphics[width = 0.45\textwidth]{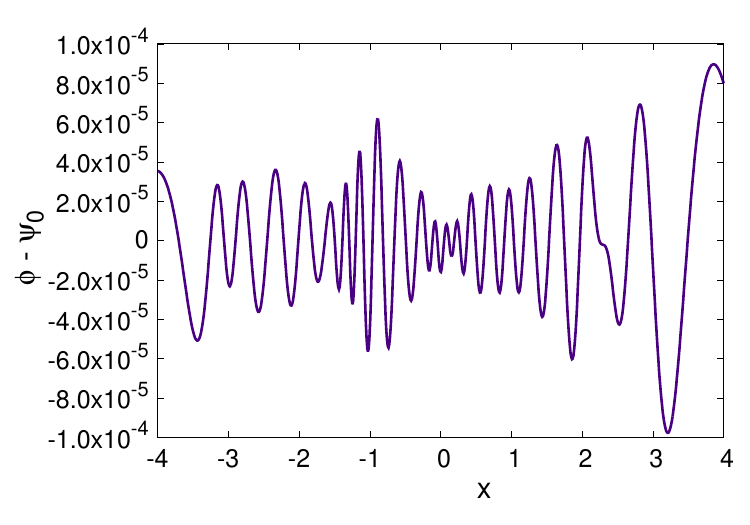}}\quad
\subfloat[]{\includegraphics[width = 0.45\textwidth]{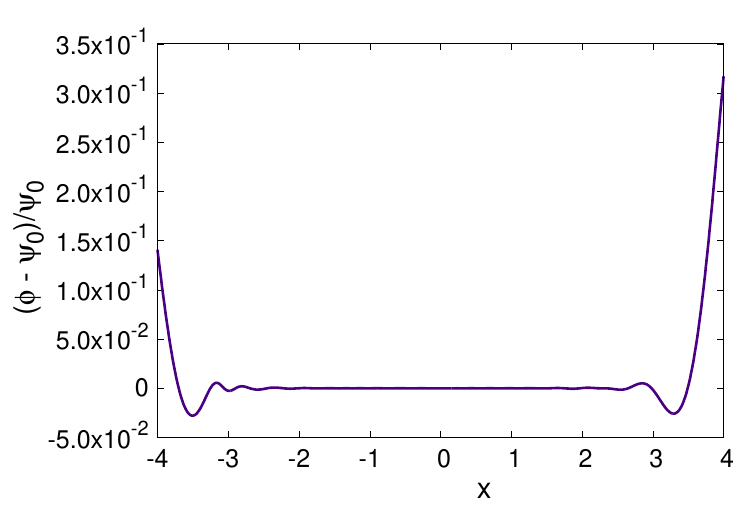}}
\caption{Difference between the DFNN approximation  and the exact solution for the ground-state of the harmonic oscillator. In panel (a) is shown the difference, whereas in panel (b)  is shown the relative difference respect to the exact solution.}
\label{harmonicPsi}
\end{figure}

\item{Linear harmonic potential:}
\begin{align}
V(x) &= \frac{1}{2}x^2 - \lambda x, \quad E_0 = \frac{1}{2}\ta 1-\lambda^2\tc \quad \psi_0 =  \frac{e^{-\frac{(x-\lambda)^2}{2}}}{\pi^{\frac{1}{4}}}. \label{linHarmOsc} 
\end{align}
We fix $\lambda = \frac{1}{2}$, and we consider the same network structure as for the harmonic oscillator. We find
\[
\bar{E}-E_0 =   10^{-6},\quad \quad |\braket{\phi|\phi} - 1| <  10^{-7}, \quad \quad \mathcal{G}(10^5) \simeq  10^{-5}.
\]
In figure \ref{linharmonicPsi} it is shown the difference between the DFNN approximation and the exact solution as a function of $x$. The agreement is satisfactory, for both the absolute and relative difference. Also in this case, at the boundary the relative difference increases as a consequence of the vanishing wave function.
The integration region $\Omega$ assumed by the essential condition \ref{ess1} is chosen to be $\Omega = [-4,4]$.

\begin{figure}
\centering
\subfloat[]{\includegraphics[width = 0.45\textwidth]{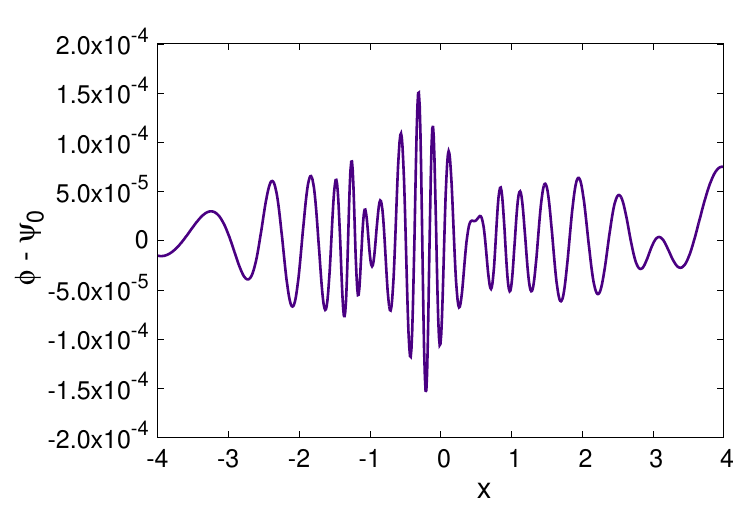}}\quad
\subfloat[]{\includegraphics[width = 0.45\textwidth]{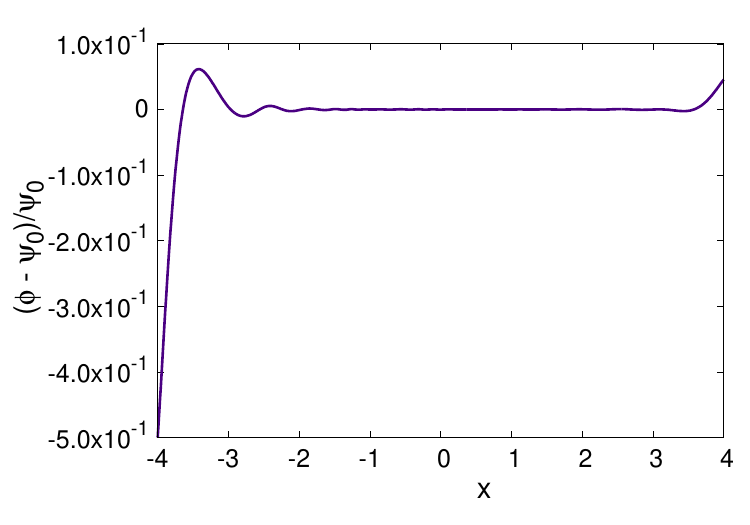}}
\caption{Difference between the DFNN approximation  and the exact solution for the ground-state of the linear harmonic oscillator of Eq.~\eqref{linHarmOsc}. In panel (a)  is shown the  difference, whereas in panel (b)  is shown the relative difference respect to the exact solution.}
\label{linharmonicPsi}
\end{figure}

\item{Quartic harmonic potential:}
\begin{align}
V(x) &= \frac{1}{2}x^2 + \lambda x^4, \quad E_0, \psi_0  \ \text{ known only in perturbation theory}.
\label{QuarticPot}
\end{align}
For $\lambda^2\ll \lambda$,  the ground-state energy gets the following correction:
\[
E^{(1)}_0 = E_0^{ho} + \braket{\psi_0|  \lambda x^4 |\psi_0} = E_0^{ho} + \frac{3\lambda}{4},
\]
where $E_0^{ho}$ is the ground-state energy of the pure harmonic oscillator.
With the same DFNN structure as before, we obtain for a value of $\lambda = 0.01$
\[
\bar{E} - E^{(1)}_0 \simeq 10^{-4}, \quad |\braket{\phi|\phi} - 1| <  10^{-7}, \quad \quad \mathcal{G}(10^5) \simeq  10^{-5},
\]
where the $\Omega$ region is fixed to $\Omega=[-4,4]$.
The discrepancy between $\bar{E}$ and $E^{(1)}_0$ is fully consistent with the expectation that the higher-order corrections are proportional to $\lambda^2 = 10^{-4}$.

We considered also the case of `large' non-perturbative $\lambda =1$. In order to achieve a good convergence in this case we change the structure of the network to a DFNN with six hidden layers, where the first two layers have twenty nodes each and the remaining four layers have ten nodes each. We also restrict the integration region to be $\Omega=[-3,3]$. We obtain:
\[
\bar{E} = 0.803771,\quad |\braket{\phi|\phi} - 1| \simeq  10^{-7}, \quad \quad \mathcal{G}(10^5) \simeq  10^{-5}.
\]
In figure \ref{quarticPsi} are shown the results for the two values of $\lambda$ in comparison to the harmonic oscillator wave function. 
As expected, for $\lambda=0.01$ the approximate solution is quite well reproduced by the harmonic-oscillator wave function, whereas for $\lambda=1$ the approximate solution is more localized near the origin with respect to to the harmonic-oscillator solution, since the anharmonic contribution in the potential has a faster increase with $x$ than the harmonic term. 

\begin{figure}
\centering
\subfloat[]{\includegraphics[width = 0.45\textwidth]{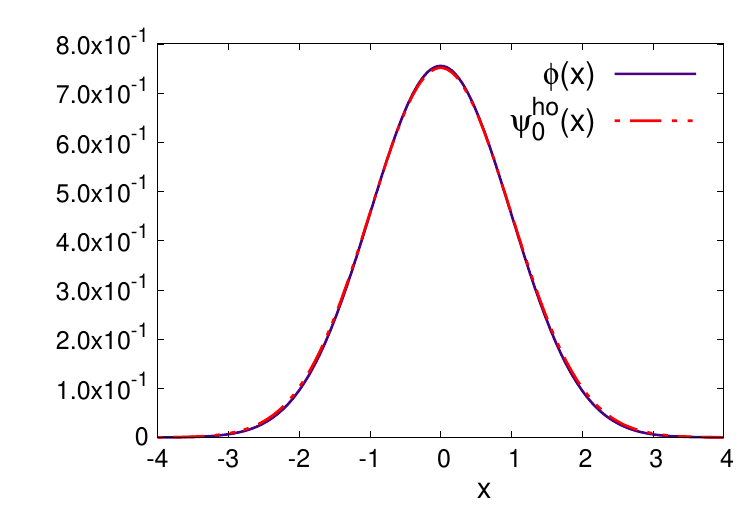}}\quad
\subfloat[]{\includegraphics[width = 0.45\textwidth]{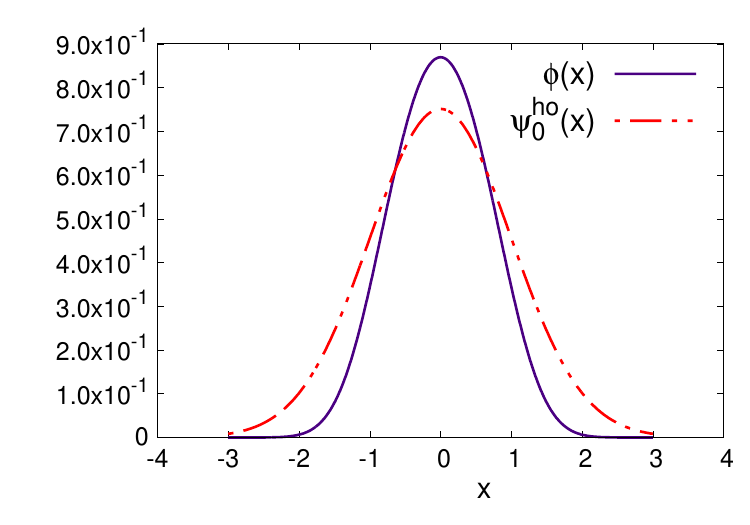}}
\caption{The results of the DFNN approximation $\phi(x)$ for the quartic oscillator for $\lambda=0.01$ (panel (a)) and for $\lambda=1$ (panel (b)) compared to the exact solution for the harmonic oscillator $\psi_0^{ho}$. The solid purple curve is the DFNN approximated solution for the quartic potential in Eq.~\eqref{QuarticPot}, whereas the red dashed-dot curve is the analytical solution for the harmonic oscillator.}
\label{quarticPsi}
\end{figure}

\item{Exponential potential:}
\begin{align}
V(x) &= e^{|x|}, \quad E_0, \psi_0  \ \text{ unknown}.
\label{expPot}
\end{align}
This potential is an example of a full-confining potential, like the harmonic potential, but without  the simple polynomial functional form.
In order to obtain the desired approximation, we use a DFNN with six hidden layers where the first two layers have twenty nodes each and the remaining four layers have ten nodes each. The integration region is set to $\Omega=[-3,3]$. 
We obtain
\[
\bar{E} = 2.060040,\quad |\braket{\phi|\phi} - 1| \simeq  2\times 10^{-6}, \quad \quad \mathcal{G}(10^5) \simeq  10^{-5}.
\]
The DFNN approximation of the ground-state wave function for the exponential potential is shown in figure \ref{expPotFigure}.
\begin{figure}
\centering
\includegraphics[width = 0.45\textwidth]{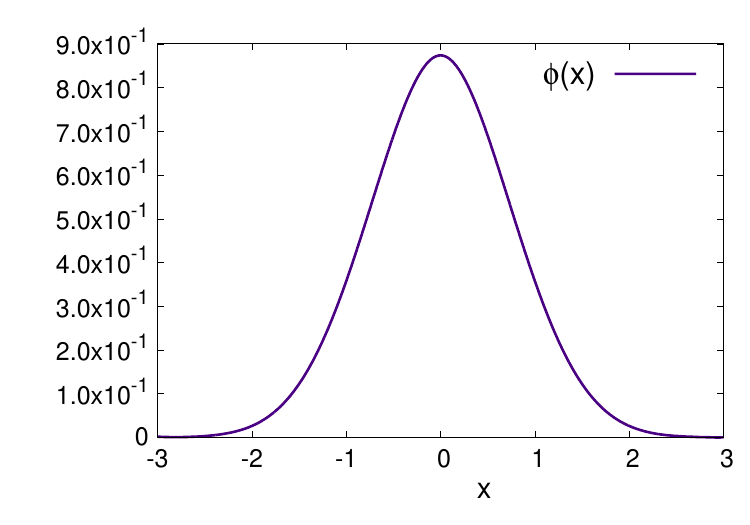}
\caption{The DFNN approximation of the ground-state wave function for the exponential potential in Eq.~\eqref{expPot}. }
\label{expPotFigure}
\end{figure}

\item{Morse potential:}
\begin{align}
V(x) &= e^{-2x}-2e^{-x}+1,\quad E_0 = 1-\ta\frac{\sqrt{2}-2}{2}\tc^2, \quad \psi_0(x) =  \ta 2\sqrt{2}e^{-x}\tc^{\sqrt{2}-1/2} e^{-e^{-x}/2}.
\label{MorsePotential}
\end{align}

\begin{figure}
\centering
\includegraphics[width = 0.45\textwidth]{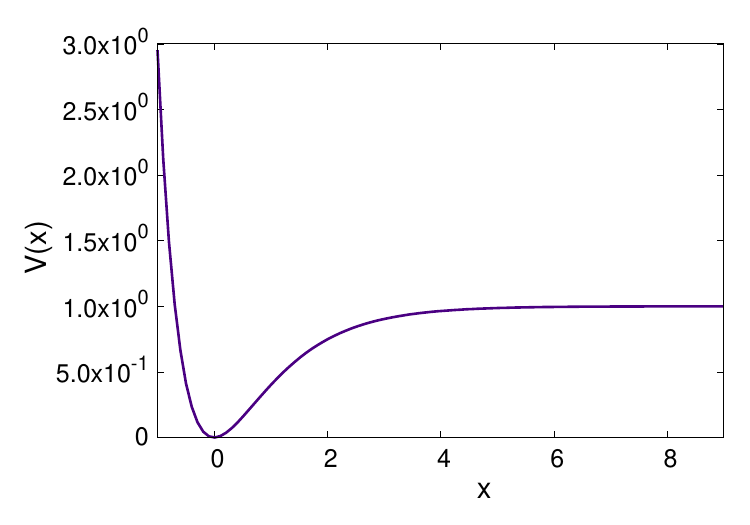}
\caption{Morse potential of Eq.~\eqref{MorsePotential}. }
\label{MorsePotFigure}
\end{figure}
This potential differs from all the previous full-confining potentials: it rises without constraints for large and negative values of $x$, whereas for large and positive values of $x$ it saturates  to a constant value (for the version in Eq.~\eqref{MorsePotential} the saturation value is $1$),  as shown in figure \ref{MorsePotFigure}.

Using a DFNN with six hidden layers with twenty nodes each, we obtain:
\[
\bar{E}-E_0 =  2\times 10^{-6},\quad \quad |\braket{\phi|\phi} - 1| <  10^{-7}, \quad \quad \mathcal{G}(10^5) \simeq  10^{-5}.
\]
In figure \ref{morsePsi} are shown the numerical solution and the difference between the analytical and numerical solutions. 
\begin{figure}
\centering
\subfloat[]{\includegraphics[width = 0.45\textwidth]{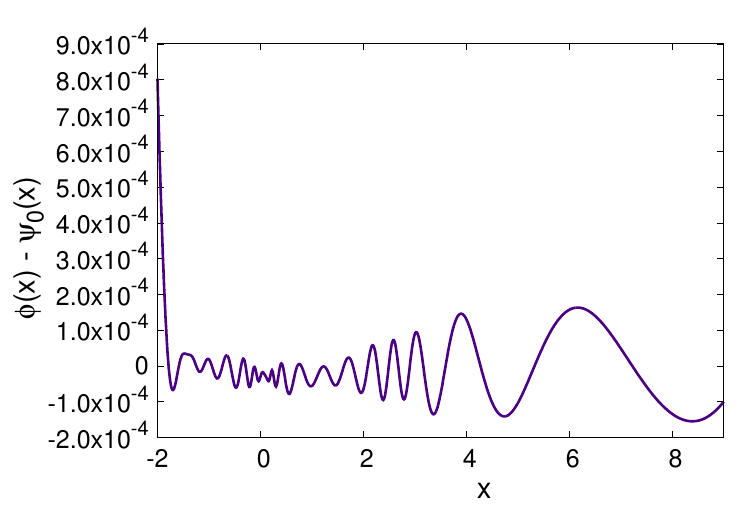}}\quad
\subfloat[]{\includegraphics[width = 0.45\textwidth]{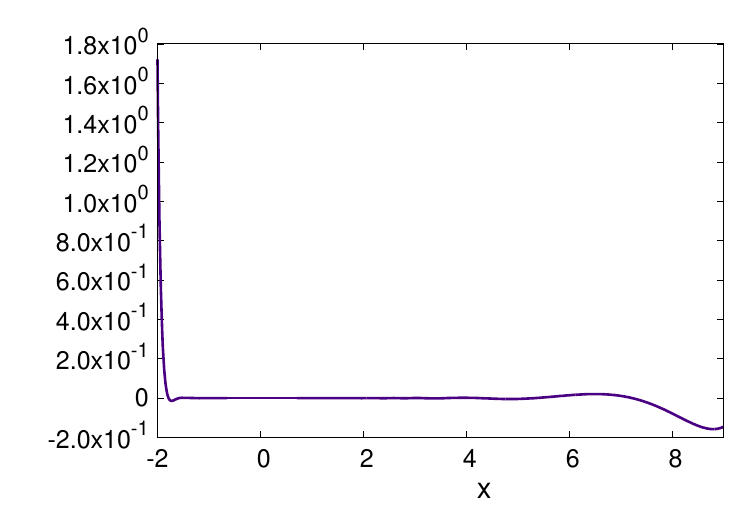}}\\
\subfloat[]{\includegraphics[width = 0.45\textwidth]{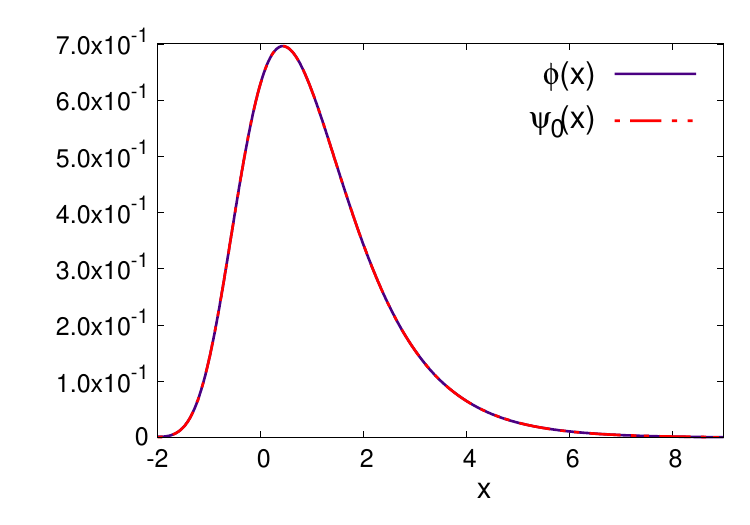}}
\caption{Difference between the DFNN approximation  and the exact solution for the ground-state of the Morse potential of Eq.~\eqref{MorsePotential}. In panel (a)  is shown the absolute difference, whereas in panel (b)  is shown the relative difference respect to the exact solution. In panel (c) the two solutions are shown together.}
\label{morsePsi}
\end{figure}
The region $\Omega$ is chosen to be $\Omega = [-2,9]$.

\end{itemize}

\section{Generalization to multiple dimensions and execution time considerations}
\label{SecGeneralization}
We presented the results for the simple case of a mono-dimensional system, in which the integrations can be performed efficiently using a grid for the input-space domain. 
For a multi-dimensional generalization, albeit the overall logic of the variational method remains the same, one has to use either a fast meshed integration algorithm or a Monte Carlo based approach. 
As highlighted in \cite{Sirignano_2018}, for a large dimension of the input space the execution time required to compute the first- and second-order derivatives becomes a limiting factor. Also the number of outputs influences the overall execution time.

To partially compensate the loss of speed for higher dimensional inputs, one could use different approaches.
First, one can reduce the order of the derivatives in the variational problems by using integration by parts and by cleverly rewriting the energy functional only in terms of first derivatives of the DFNN approximation, see \cite{kharazmi2019variational}.
Second, if only the diagonal of the hessian of the inputs is needed, one can modify the algorithm to compute only the relevant double derivatives, decreasing both memory consumption and evaluation time. 
Third, one can implement a parallelization of the evaluation of the variational integrals. 
A combination of all the previous strategies can effectively reduce the evaluation time for the overall variational problem.

\begin{figure}
\centering
\subfloat[]{\includegraphics[width = 0.45\textwidth]{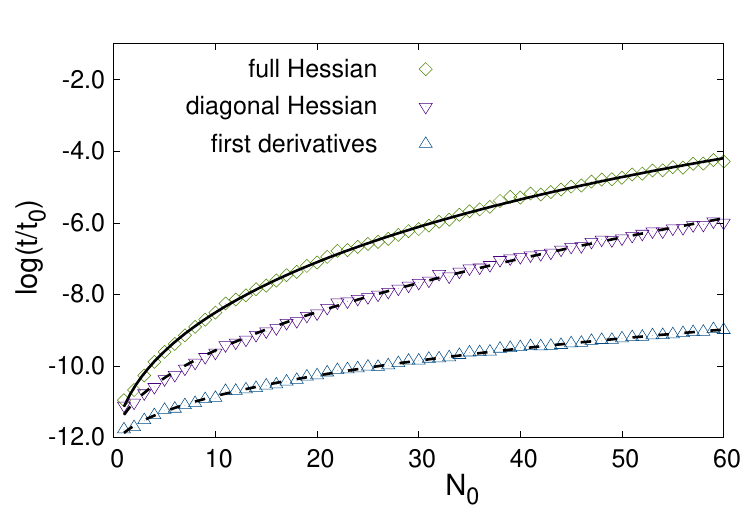}}\quad
\subfloat[]{\includegraphics[width = 0.45\textwidth]{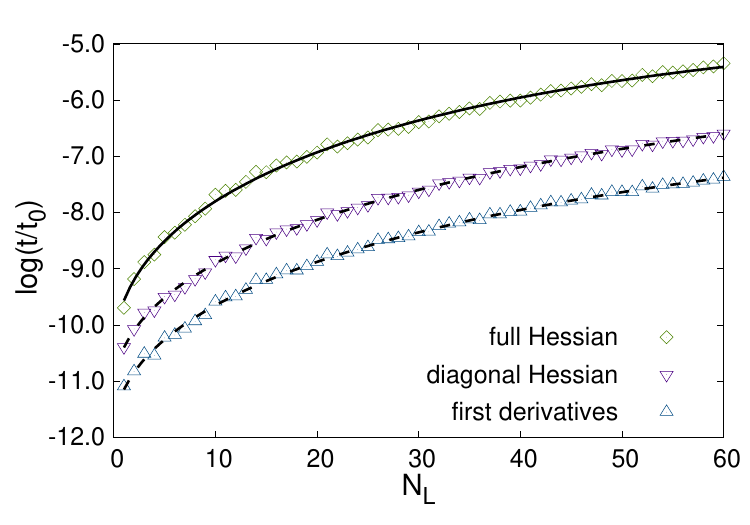}}
\caption{Logarithmic of the evaluation time $\log(t/t_0)$ with $t_0=1$ s of a DFNN as a function of the input-space size (panel (a)) and as a function of the output-space size (panel (b)).
In both cases the DFNN has five hidden layers with ten nodes each. For panel (a) the DFNN has a single output, for panel (b) the DFNN has four inputs.}
\label{speed}
\end{figure}

To better highlight the behavior of our implementation of the algorithm of section \ref{SecAnalytDer}, we study the time required for a single evaluation of a DFNN with five hidden layers of ten nodes each in two cases: variable number of inputs $N_0$, one output $N_L=1$; variable number of outputs $N_L$, four inputs $N_0=4$.
In panel (a) of figure \ref{speed}  is displayed the logarithm of the evaluation time $\log(t/t_0)$, with $t_0= 1$ s (averaged over $200$ iterations) in the case of variable number of inputs. In panel (b) of figure \ref{speed}  is displayed the logarithm of the evaluation time $\log(t/t_0)$, with $t_0= 1$ s (averaged over $200$ iterations) in the case of variable number of outputs.

The two panels in figure \ref{speed} display also a trend curve obtained by a fit of the logarithm of the measured times with the functional form
\[
\log(t(x)/t_0) \simeq f(x) = ax+b\sqrt{x}+c, \quad \quad x=N_0 \text{ or } x=N_L.
\label{trendLineEq}
\]
The parameters for the different cases shown in figure \ref{speed} are reported in table \ref{speed_par}. The most notable effect is in the case of large number of inputs, for which the first derivative approach is approximately two order of magnitudes faster then the full Hessian approach and one order of magnitude faster than the diagonal Hessian approach.

\begin{table}
\centering
\renewcommand{\arraystretch}{1.2}
\begin{tabular}{c|c|c|c|c|c|c}
\multirow{2}{2cm}{Parameters} & \multicolumn{3}{c|}{$N_0$} & \multicolumn{3}{c}{$N_L$} \\
\cline{2-7}
& Full Hess. & Diag. Hess. & First der. & Full Hess. & Diag. Hess. & First der. \\
\hline
$a$ & $-0.041$ & $-0.004$     & $-0.010$ & $-0.044$ & $-0.028$ & $-0.029$ \\
$b$ & $1.386$ & $0.848$       & $0.518$ & $0.998$ & $0.806$ & $0.813$ \\
$c$ & $-12.483$ & $-12.210$ & $-12.381$ & $-10.519$ & $-11.177$ & $-11.931$ \\
\end{tabular}
\caption{Parameters of Eq.~\eqref{trendLineEq} for the trend lines displayed in figure \ref{speed}. The columns two, three and four refer to variable number of inputs of the DFNN in the case of the full Hessian calculation or only the diagonal of the Hessian or just the first derivatives, respectively. The columns five six and seven refer to the same calculations ion the case of variable number of outputs. }
\label{speed_par}
\end{table}

\section{Conclusions}
\label{SecConcl}
In this work, we presented the full recursive algorithm that allows the computation of the first- and second-order derivatives of an arbitrary deep feed forward neural network, along with the gradient with respect to the weights and biases. 
In sections \ref{sectQM} and \ref{SecExamples}, we used the algorithm, presented in section \ref{SecAnalytDer}, to study the eigenvalue variational problem in quantum mechanics, using  various mono-dimensional potentials as specific examples. 
In section \ref{SecGeneralization}, we briefly discussed the possible extensions to multidimensional problems, analyzing the evaluation time of the algorithm as a function of the input and output space size.
With a fast implementation of the matrix algebra and with some manipulation of the formulation of the variational problem, one can generalize the examples to multi-dimensional potential. 
A fast implementation of the algorithm opens the possibility of accurately study a variety of systems, like quantum many-body systems and, possibly, quantum-field theory systems. The latter are especially interesting for the possibility of studying the hadrons wave functions using DFNNs, which could be connected to the extraction of non-perturbative information in the form of parton distribution functions, as already performed using neural networks. \cite{forte2020parton}. 
\section*{Acknowledgments}
The author wish to thank Barbara Pasquini, Oreste Nicrosini, Valerio Bertone and Rabah Abdul Khalek for useful discussions.
The work of S.R.~is part of a project that has received funding from the European Union's Horizon 2020 research and innovation programme under grant agreement STRONG - 2020 - No~824093.
\section*{Code}
A C implementation of the algorithm with the proposed examples is publicly available, under a MIT license, at the github repository \cite{Rodini:2021}.

\biboptions{sort&compress}
\bibliographystyle{model6-num-names}

\begin{thebibliography}{21}
\providecommand{\natexlab}[1]{#1}
\providecommand{\url}[1]{\texttt{#1}}
\providecommand{\href}[2]{#2}
\providecommand{\path}[1]{#1}
\providecommand{\DOIprefix}{doi:}
\providecommand{\ArXivprefix}{arXiv:}
\providecommand{\URLprefix}{URL: }
\providecommand{\Pubmedprefix}{pmid:}
\providecommand{\doi}[1]{\href{http://dx.doi.org/#1}{\path{#1}}}
\providecommand{\Pubmed}[1]{\href{pmid:#1}{\path{#1}}}
\providecommand{\BIBand}{and}
\providecommand{\bibinfo}[2]{#2}
\ifx\xfnm\undefined \def\xfnm[#1]{\unskip,\space#1}\fi
\makeatletter\def\@biblabel#1{#1.}\makeatother
\bibitem[{G.(1989)}]{Cybenko:1989}
\bibinfo{author}{G.\xfnm[ C.]}.
\newblock \bibinfo{title}{{Approximation by Superposition of a Sigmoidal
  Function}}.
\newblock \emph{\bibinfo{journal}{Math Control Signals Systems}}
  \bibinfo{year}{1989};\bibinfo{volume}{2}:\bibinfo{pages}{303--314}.
\newblock \DOIprefix\doi{10.1007/BF02551274}.
\bibitem[{Avrutskiy(2020)}]{Avrutskiy_2020}
\bibinfo{author}{Avrutskiy\xfnm[ V.]}.
\newblock \bibinfo{title}{Neural networks catching up with finite differences
  in solving partial differential equations in higher dimensions}.
\newblock \emph{\bibinfo{journal}{Neural Computing and Applications}}
  \bibinfo{year}{2020};\bibinfo{volume}{32}.
\newblock \DOIprefix\doi{10.1007/s00521-020-04743-8}.
\bibitem[{Dockhorn(2019)}]{dockhorn2019discussion}
\bibinfo{author}{Dockhorn\xfnm[ T.]}.
\newblock \bibinfo{title}{A discussion on solving partial differential
  equations using neural networks}.
\newblock \bibinfo{year}{2019}.
\newblock \href{http://arxiv.org/abs/1904.07200}{\tt arXiv:1904.07200}.
\bibitem[{Lee and Kang(1990)}]{LEE1990110}
\bibinfo{author}{Lee\xfnm[ H.]}, \bibinfo{author}{Kang\xfnm[ I.S.]}.
\newblock \bibinfo{title}{Neural algorithm for solving differential equations}.
\newblock \emph{\bibinfo{journal}{Journal of Computational Physics}}
  \bibinfo{year}{1990};\bibinfo{volume}{91}(\bibinfo{number}{1}):\bibinfo{pages}{110--131}.
\newblock \URLprefix
  \url{https://www.sciencedirect.com/science/article/pii/002199919090007N}.
  \DOIprefix\doi{https://doi.org/10.1016/0021-9991(90)90007-N}.
\bibitem[{Raissi et~al.(2017{\natexlab{a}})Raissi, Perdikaris and
  Karniadakis}]{raissi2017physics1}
\bibinfo{author}{Raissi\xfnm[ M.]}, \bibinfo{author}{Perdikaris\xfnm[ P.]},
  \bibinfo{author}{Karniadakis\xfnm[ G.E.]}.
\newblock \bibinfo{title}{{Physics Informed Deep Learning (Part I): Data-driven
  Solutions of Nonlinear Partial Differential Equations}}.
\newblock \bibinfo{year}{2017}{\natexlab{a}}.
\newblock \href{http://arxiv.org/abs/1711.10561}{\tt arXiv:1711.10561}.
\bibitem[{Raissi et~al.(2017{\natexlab{b}})Raissi, Perdikaris and
  Karniadakis}]{raissi2017physics2}
\bibinfo{author}{Raissi\xfnm[ M.]}, \bibinfo{author}{Perdikaris\xfnm[ P.]},
  \bibinfo{author}{Karniadakis\xfnm[ G.E.]}.
\newblock \bibinfo{title}{{Physics Informed Deep Learning (Part II):
  Data-driven Discovery of Nonlinear Partial Differential Equations}}.
\newblock \bibinfo{year}{2017}{\natexlab{b}}.
\newblock \href{http://arxiv.org/abs/1711.10566}{\tt arXiv:1711.10566}.
\bibitem[{Lagaris et~al.(1998)Lagaris, Likas and Fotiadis}]{Lagaris_1998}
\bibinfo{author}{Lagaris\xfnm[ I.]}, \bibinfo{author}{Likas\xfnm[ A.]},
  \bibinfo{author}{Fotiadis\xfnm[ D.]}.
\newblock \bibinfo{title}{{Artificial neural networks for solving ordinary and
  partial differential equations}}.
\newblock \emph{\bibinfo{journal}{IEEE Transactions on Neural Networks}}
  \bibinfo{year}{1998};\bibinfo{volume}{9}(\bibinfo{number}{5}):\bibinfo{pages}{987?1000}.
\newblock \URLprefix \url{http://dx.doi.org/10.1109/72.712178}.
  \DOIprefix\doi{10.1109/72.712178}.
\bibitem[{Lagaris et~al.(2000)Lagaris, Likas and Papageorgiou}]{Lagaris_2000}
\bibinfo{author}{Lagaris\xfnm[ I.]}, \bibinfo{author}{Likas\xfnm[ A.]},
  \bibinfo{author}{Papageorgiou\xfnm[ D.]}.
\newblock \bibinfo{title}{Neural-network methods for boundary value problems
  with irregular boundaries}.
\newblock \emph{\bibinfo{journal}{IEEE transactions on neural networks / a
  publication of the IEEE Neural Networks Council}}
  \bibinfo{year}{2000};\bibinfo{volume}{11}:\bibinfo{pages}{1041--9}.
\newblock \DOIprefix\doi{10.1109/72.870037}.
\bibitem[{Lagaris et~al.(1997)Lagaris, Likas and Fotiadis}]{Lagaris_1997}
\bibinfo{author}{Lagaris\xfnm[ I.]}, \bibinfo{author}{Likas\xfnm[ A.]},
  \bibinfo{author}{Fotiadis\xfnm[ D.]}.
\newblock \bibinfo{title}{{Artificial neural network methods in quantum
  mechanics}}.
\newblock \emph{\bibinfo{journal}{Computer Physics Communications}}
  \bibinfo{year}{1997};\bibinfo{volume}{104}(\bibinfo{number}{1-3}):\bibinfo{pages}{1?14}.
\newblock \URLprefix \url{http://dx.doi.org/10.1016/S0010-4655(97)00054-4}.
  \DOIprefix\doi{10.1016/s0010-4655(97)00054-4}.
\bibitem[{Guidetti et~al.(2021)Guidetti, Muia, Welling and
  Westphal}]{Guidetti:2021xvb}
\bibinfo{author}{Guidetti\xfnm[ V.]}, \bibinfo{author}{Muia\xfnm[ F.]},
  \bibinfo{author}{Welling\xfnm[ Y.]}, \bibinfo{author}{Westphal\xfnm[ A.]}.
\newblock \bibinfo{title}{{dNNsolve: an efficient NN-based PDE solver}}
  \bibinfo{year}{2021};\href{http://arxiv.org/abs/2103.08662}{\tt
  arXiv:2103.08662}.
\bibitem[{Kharazmi et~al.(2019)Kharazmi, Zhang and
  Karniadakis}]{kharazmi2019variational}
\bibinfo{author}{Kharazmi\xfnm[ E.]}, \bibinfo{author}{Zhang\xfnm[ Z.]},
  \bibinfo{author}{Karniadakis\xfnm[ G.E.]}.
\newblock \bibinfo{title}{Variational physics-informed neural networks for
  solving partial differential equations}.
\newblock \bibinfo{year}{2019}.
\newblock \href{http://arxiv.org/abs/1912.00873}{\tt arXiv:1912.00873}.
\bibitem[{E(2018)}]{Wu_2017}
\bibinfo{author}{E W.\xfnm[ Y.B.]}.
\newblock \bibinfo{title}{A deep learning-based numerical algorithm for solving
  variational problems}.
\newblock \emph{\bibinfo{journal}{Commun Math Stat}}
  \bibinfo{year}{2018};\bibinfo{volume}{6}:\bibinfo{pages}{1--12}.
\newblock \DOIprefix\doi{10.1007/s40304-018-0127-z}.
\bibitem[{Nakanishi and Sugawara(2000)}]{nakanishi2000}
\bibinfo{author}{Nakanishi\xfnm[ H.]}, \bibinfo{author}{Sugawara\xfnm[ M.]}.
\newblock \bibinfo{title}{Numerical solution of the schroedinger equation by a
  microgenetic algorithm}.
\newblock \emph{\bibinfo{journal}{Chemical Physics Letters}}
  \bibinfo{year}{2000};\bibinfo{volume}{327}:\bibinfo{pages}{429--438}.
\newblock \DOIprefix\doi{10.1016/S0009-2614(00)00913-1}.
\bibitem[{Sugawara(2001)}]{sugawara2001}
\bibinfo{author}{Sugawara\xfnm[ M.]}.
\newblock \bibinfo{title}{{Numerical solution of the Schr\"odinger equation by
  neural network and genetic algorithm}}.
\newblock \emph{\bibinfo{journal}{Computer Physics Communications}}
  \bibinfo{year}{2001};\bibinfo{volume}{140}:\bibinfo{pages}{366--380}.
\newblock \DOIprefix\doi{10.1016/S0010-4655(01)00286-7}.
\bibitem[{Marim et~al.(2003)Marim, Lemes and {Dal Pino}}]{MARIM2003159}
\bibinfo{author}{Marim\xfnm[ L.]}, \bibinfo{author}{Lemes\xfnm[ M.]},
  \bibinfo{author}{{Dal Pino}\xfnm[ A.]}.
\newblock \bibinfo{title}{Ground-state of silicon clusters by neural network
  assisted genetic algorithm}.
\newblock \emph{\bibinfo{journal}{Journal of Molecular Structure: THEOCHEM}}
  \bibinfo{year}{2003};\bibinfo{volume}{663}(\bibinfo{number}{1}):\bibinfo{pages}{159--165}.
\newblock \URLprefix
  \url{https://www.sciencedirect.com/science/article/pii/S0166128003007553}.
  \DOIprefix\doi{https://doi.org/10.1016/j.theochem.2003.08.123}.
\bibitem[{Sirignano and Spiliopoulos(2018)}]{Sirignano_2018}
\bibinfo{author}{Sirignano\xfnm[ J.]}, \bibinfo{author}{Spiliopoulos\xfnm[
  K.]}.
\newblock \bibinfo{title}{Dgm: A deep learning algorithm for solving partial
  differential equations}.
\newblock \emph{\bibinfo{journal}{Journal of Computational Physics}}
  \bibinfo{year}{2018};\bibinfo{volume}{375}:\bibinfo{pages}{1339--1364}.
\newblock \URLprefix \url{http://dx.doi.org/10.1016/j.jcp.2018.08.029}.
  \DOIprefix\doi{10.1016/j.jcp.2018.08.029}.
\bibitem[{Abdul~Khalek and Bertone(2020)}]{AbdulKhalek:2020uza}
\bibinfo{author}{Abdul~Khalek\xfnm[ R.]}, \bibinfo{author}{Bertone\xfnm[ V.]}.
\newblock \bibinfo{title}{{On the derivatives of feed-forward neural networks}}
  \bibinfo{year}{2020};\href{http://arxiv.org/abs/2005.07039}{\tt
  arXiv:2005.07039}.
\bibitem[{Sakurai and Napolitano(2020)}]{Sakurai:2011zz}
\bibinfo{author}{Sakurai\xfnm[ J.J.]}, \bibinfo{author}{Napolitano\xfnm[ J.]}.
\newblock \bibinfo{title}{{Modern Quantum Mechanics}}.
\newblock Quantum physics, quantum information and quantum computation;
  \bibinfo{publisher}{Cambridge University Press}; \bibinfo{year}{2020}.
\newblock ISBN \bibinfo{isbn}{978-0-8053-8291-4, 978-1-108-52742-2,
  978-1-108-58728-0}.
\newblock \DOIprefix\doi{10.1017/9781108587280}.
\bibitem[{Kingma and Ba(2014)}]{Kingma:2014vow}
\bibinfo{author}{Kingma\xfnm[ D.P.]}, \bibinfo{author}{Ba\xfnm[ J.]}.
\newblock \bibinfo{title}{{Adam: A Method for Stochastic Optimization}}
  \bibinfo{year}{2014};\href{http://arxiv.org/abs/1412.6980}{\tt
  arXiv:1412.6980}.
\bibitem[{Forte and Carrazza(2020)}]{forte2020parton}
\bibinfo{author}{Forte\xfnm[ S.]}, \bibinfo{author}{Carrazza\xfnm[ S.]}.
\newblock \bibinfo{title}{Parton distribution functions}.
\newblock \bibinfo{year}{2020}.
\newblock \href{http://arxiv.org/abs/2008.12305}{\tt arXiv:2008.12305}.
\bibitem[{Rodini(2021)}]{Rodini:2021}
\bibinfo{author}{Rodini\xfnm[ S.]}.
\newblock \bibinfo{title}{nn{D}{E}: a library for first and second order neural
  network derivatives}.
\newblock \bibinfo{year}{2021}.
\newblock \URLprefix \url{https://github.com/slrodini/nnDE_public}.

\end{thebibliography}

\end{document}